\documentclass[aps,prl,twocolumn,groupedaddress,showpacs,superscriptaddress,floatfix]{revtex4-1}
\usepackage{pgf}
\usepackage{amsmath}
\usepackage{color}
\usepackage[colorlinks,allcolors=blue,hypertexnames=false]{hyperref}

\newcommand{\refsubfig}[2]{\hyperref[#1]{\ref*{#1}(#2)}}
\renewcommand{\eqref}[1]{\hyperref[#1]{(\ref*{#1})}}
\begin{document}
	\title{Universal Scaling Laws for Dense Particle Suspensions in Turbulent Wall-Bounded Flows}
	\author{Pedro Costa}
	\email[]{p.simoescosta@tudelft.nl}
	\affiliation{Laboratory for Aero and Hydrodynamics, Delft University of Technology, Leeghwaterstraat 21, NL-2628 CA Delft, The Netherlands}
	\author{Francesco Picano}
	\affiliation{Department of Industrial Engineering, University of Padova, Via Venezia 1, 35131 Padua, Italy}
	\author{Luca Brandt}
	\affiliation{SeRC (Swedish e-Science Research Centre) and Linn\'{e} FLOW Centre, KTH Mechanics, SE-100 44 Stockholm, Sweden}
	\author{Wim-Paul Breugem}
	\affiliation{Laboratory for Aero and Hydrodynamics, Delft University of Technology, Leeghwaterstraat 21, NL-2628 CA Delft, The Netherlands}
	\date{\today}
\begin{abstract}
The macroscopic behavior of dense suspensions of neutrally-buoyant spheres in turbulent plane channel flow is examined. We show that particles larger than the smallest turbulence scales cause the suspension to deviate from the continuum limit in which its dynamics is well described by an effective suspension viscosity. This deviation is caused by the formation of a particle layer close to the wall with significant slip velocity. By assuming two distinct transport mechanisms in the near-wall layer and the turbulence in the bulk, we define an effective wall location such that the flow in the bulk can still be accurately described by an effective suspension viscosity. We thus propose scaling laws for the mean velocity profile of the suspension flow, together with a master equation able to predict the increase in drag as function of the particle size and volume fraction. 
\end{abstract}
\maketitle
%
%%fakesection %INTRODUCTION
%
Turbulent, wall-bounded suspensions appear widely in environmental and industrial contexts. These suspensions are often dense, i.e.\ the volume fraction is sufficiently high that particle-particle and particle-fluid interactions strongly influence the macroscopic flow dynamics. In many cases, the suspended particles have a \emph{finite size} -- comparable to or larger than the smallest scales in the flow, and particle inertia plays an important role \cite{Balachandar-and-Eaton-ARFM-2010}. 

The flow of suspensions under laminar conditions has been thoroughly studied since Einstein~\cite{Einstein1906} analytically derived an expression for  the effective viscosity of a suspension of rigid spheres in the dilute and viscous limit: $\nu^e/\nu =  1+(5/2)\Phi$, where $\nu$ is the kinematic viscosity of the suspending fluid, and $\Phi$ the bulk solid volume fraction. In dense cases, the rheology of laminar suspensions is usually characterized by semi-empirical formulas for the effective viscosity~\cite{Stickel-and-Powell-ARFM-2005,Guazzelli2011}. 

When the Reynolds number (which quantifies the importance of fluid inertial to viscous effects) is sufficiently high, the flow becomes turbulent, exhibiting chaotic and multiscale dynamics. Wall-bounded turbulent flows are characterized by at least one inhomogeneous direction and by the constraint of vanishing velocity at the wall, which makes their analysis even more complicated. 
For simplicity, we consider the canonical case of a pressure-driven turbulent plane-channel flow laden with neutrally-buoyant particles, defined by the bulk Reynolds number $\mathrm{Re}_b=U_b\, 2h/\nu$, where $U_b$ is the bulk velocity (i.e.\ averaged over the entire domain) and $h$ the half channel height. 
In the single-phase limit, the most well-known results from classical turbulence theory are the scaling laws for the mean velocity and the associated drag, or pressure loss. This is obtained by dividing the flow into two regions: the \emph{inner layer}, close to the wall, $y \ll h$, with relevant velocity and length scales $u_\tau$ and $\delta_v$, and the \emph{outer layer}, away from the wall, $y\gg\delta_v$, governed by $u_\tau$ and $h$; here $u_\tau = \sqrt{\tau_w/\rho}$ is the friction velocity, $\tau_w$ the wall shear stress, $\delta_v=\nu/u_\tau$ the viscous wall unit and $\rho$ the fluid mass density. 
%
%Luca, I prefer to say that the inner layer is away from the core, and not close to the wall, because thats its definition (you can be away from the wall but still in the inner layer).

At high-enough friction Reynolds number, $\mathrm{Re}_\tau=h/\delta_v=u_\tau h/\nu \gtrsim 100$, corresponding to $\mathrm{Re}_b \gtrsim 3000$~\cite{Pope2000}, an \emph{overlap region} exists, $\delta_v \ll y \ll h$.  Here a logarithmic law can be derived for the inner-scaled mean velocity profile, $u/u_\tau = (1/\kappa)\ln(y/\delta_v) + B$, and for the outer-scaled defect law $(U_c-u)/u_\tau = -(1/\kappa)\ln(y/h) + B_d$, with $U_c$ the centerline velocity, $\kappa \approx 0.41$ the so-called von K\'arm\'an constant, $B\approx 5.2$ and $B_d \approx 0.2$.
These simple scaling laws, derived in 1930~\cite{vonKarman1930}, have been confirmed by many numerical and experimental studies (see e.g.~\cite{Smits-et-al-ARFM-2011} for a review). Their importance is unquestionable to predict the overall drag~\cite{Dean1978} and as basis for many near-wall  closure models currently used in computational fluid dynamics~\cite{Patel-et-al-AIAA-1985}.

At the very high Reynolds numbers typically encountered in practice, the suspended particles are larger than the smallest turbulent scales ($\sim \delta_v$) and the single-phase approach fails to reproduce the behavior of turbulent channel flows of dense suspensions even when accounting for an effective suspension viscosity \cite{Matas-et-al-PRL-2003,Picano-et-al-JFM-2015,Prosperetti-JFM-2015}.

In this Letter we propose scaling laws for turbulent wall-bounded suspension flows. These are characterized by three parameters: the bulk Reynolds number $\mathrm{Re}_b$, the bulk solid volume fraction $\Phi$ and the particle diameter $D_p/h$. These laws are capable of predicting the mean velocity and drag from dilute to dense cases, from large to relatively small particles and for a wide range of Reynolds numbers.\par
We use data from interface-resolved Direct Numerical Simulations (DNS). The DNS solve the Navier-Stokes equations for an incompressible Newtonian fluid in a plane channel with periodic boundary conditions in the streamwise ($x$) and spanwise ($z$) directions over lengths of $6h$ and $3h$ respectively, and no-slip and no-penetration at the bottom ($y=0$) and top ($y=2h$) walls. The flow solver is extended with an Immersed-Boundary-Method to force the fluid velocity to the local particle velocity at the particle surface~\cite{Breugem-JCP-2012}. Lubrication closures are used for short-range particle-particle and particle-wall interactions when inter-surface distances are smaller than a grid cell and a soft-sphere collision model for solid-solid contacts~\cite{Lambert-et-al-JFM-2013,Costa-et-al-PRE-2015}. The method has been tested and validated against several benchmark cases~\cite{Picano-et-al-PRL-2013,Costa-et-al-PRE-2015,Lashgari-et-al-IJMF-2016}. The flow is resolved on a uniform Cartesian grid with size $\Delta = D_p/16$. 
The computational parameters are presented in Table~\ref{tbl:params} where we also report the cases from \cite{Lashgari-et-al-PRL-2014,Picano-et-al-JFM-2015} used here for comparison. The data are complemented with an unladen single-phase reference (SPR) case at the same $\mathrm{Re}_b=12000$ and a continuum limit reference (CLR), i.e.\ the single-phase flow of a fluid with the effective viscosity $\nu^e$ of a suspension with volume fraction $\Phi=0.2$, corresponding to $\mathrm{Re}_b^e=\mathrm{Re}_b\nu/\nu^e\approx 6400$ in our case.\par
\begin{figure}
  \input{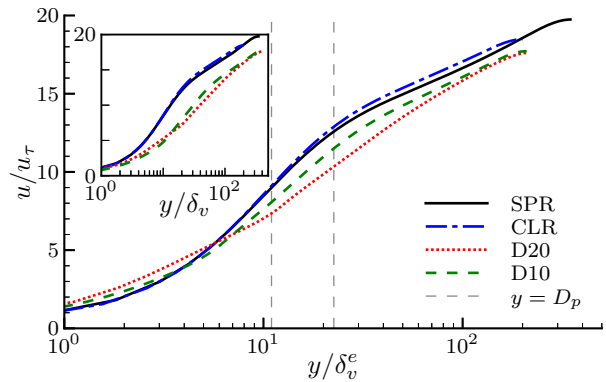}
  \caption{(color online). Mean streamwise flow velocity, $u/u_\tau$, versus the wall-normal distance in inner scaling $y/\delta_v^e$. Vertical dashed lines depict a wall-normal distance of 1 particle diameter ($y=D_p$) for cases {D10} (closest to $y=0$) and {D20}, see Table~\ref{tbl:params}. Maximum statistical error within $95\%$ confidence interval is $\pm 0.9\%$. The inset shows the same velocity profile but with the wall-normal distance scaled with $\delta_v$. \label{fig:umean_fluid_inn}}
\end{figure}

 %added paragraph
 \begin{table}%[H] add [H] placement to break table across pages
 \caption{Physical and computational parameters of the DNS database (consisting of 20 simulations). $N_p$ denotes the number of particles and $\delta_v^{sph}$ ($\gtrsim \delta_v$) the viscous wall unit for the corresponding single-phase flow at the same $\mathrm{Re}_b$.\label{tbl:params}}
 \begin{ruledtabular}
 \begin{tabular}{l r r r r r}
 Case                              & $h/D_p$ & $D_p/\delta_v^{sph}$           & $\Phi\, (\%)$ & $\mathrm{Re}_b$ & $N_p$ \\
 D10                               & $36   $ & $ 9.7                  $ & $20$ & $12\,000$           & $640\,000$ \\
 D20                               & $18   $ & $ 19.4                 $ & $20$ & $12\,000$           & $80\,000 $ \\
 D10\_2                            & $36   $ & $ 9.7                  $ & $5 $ & $12\,000$           & $160\,000$ \\
 FP \cite{Picano-et-al-JFM-2015}   & $9    $ & $ 19.9                 $ & $0-20$ &$5\,600$           & $0-10\,000$ \\
 IL \cite{Lashgari-et-al-PRL-2014} & $5    $ & $ 20.7, 32.4           $ & $0-30$ &$3\,000 , 5\,000$  & $0-2\,580$ \\
 \end{tabular}
 \end{ruledtabular}
 \end{table}
%%fakesection %EFFECTIVE VISCOSITY NOT SUFFICIENT
%
Fig.~\ref{fig:umean_fluid_inn} shows the mean velocity profile for $D_p/\delta_v \approx 10$ ({D10}) and 20 ({D20}), compared to the continuum limit reference (CLR).
The comparison between the single-phase and the two-phase flows requires a proper definition of the viscous wall unit in terms of $\nu^e$, here $\delta_v^e = \nu^e/u_\tau$. Despite the improvement with respect to the use of the classical definition of $\delta_v= \nu/u_\tau$ (see the inset of Fig.~\ref{fig:umean_fluid_inn}), the figure reveals that the particle-laden flows show a clear deviation from the classical logarithmic law. The differences with the continuum limit are for larger particles, and so is the measured increase in drag. 
The abrupt change of the slope of the profile at a wall-normal distance  of $y\sim D_p$  suggests that the deviation from the continuum limit is caused by a change in the near-wall dynamics.
%
%%fakesection %DIFFERENT NEAR-WALL DYNAMICS
%
\begin{figure}
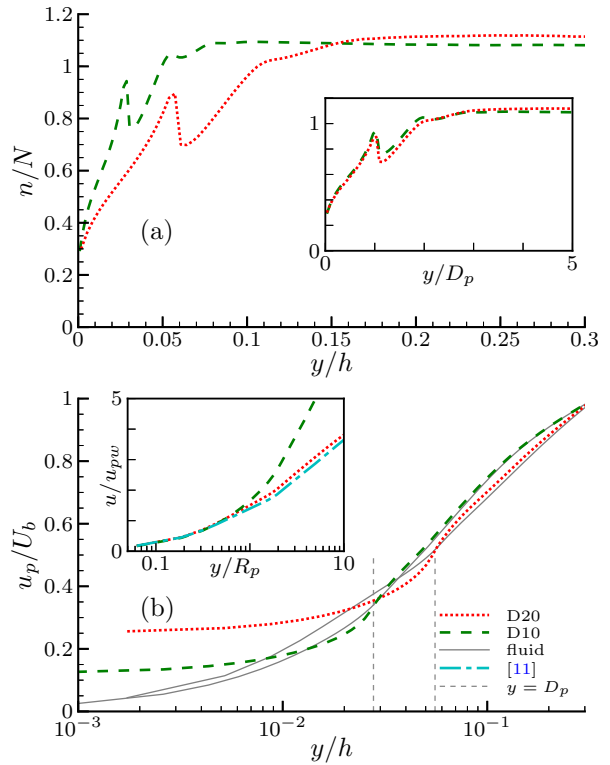

  \input{nmean_out.pgf}
  \input{umean_fluid_close_wall.pgf}
  \caption{(color online). (a) Mean particle number density $n$ divided by its bulk value $N$ versus $y/h$ in the main panel and $y/D_p$ in the inset (Maximum statistical error within $95\%$ confidence interval is $\pm 0.8\%$). (b) Mean streamwise particle and fluid velocity. The inset shows the fluid velocity, normalized with $u_{pw}$ (definition in the text) versus $y/R_p$. The data from \cite{Picano-et-al-JFM-2015} pertain to the case of $\Phi = 20\%$. Vertical dashed lines depict a wall-normal distance of 1 particle diameter ($y=D_p$) for cases {D10} (closest to $y=0$) and {D20}, see Table~\ref{tbl:params}. \label{fig:nmean_out}}
\end{figure}
Studies of laminar wall-bounded flows laden with neutrally-buoyant spheres report a structured arrangement of particles near the wall~\cite{Hampton-et-al-JR-1997,Yeo-and-Maxey-JFM-2011,Picano-et-al-PRL-2013}. This layering is attributed to the planar symmetry imposed by the wall and to stabilizing particle-particle and particle-wall interactions. Though more pronounced under laminar conditions, this phenomenon is also present in turbulent suspensions~\cite{Lashgari-et-al-PRL-2014, Picano-et-al-JFM-2015}. Fig.~\refsubfig{fig:nmean_out}{a} presents the mean local number density $n$, normalized with the corresponding bulk value $N$, for cases {D10} and {D20} (see Table~\ref{tbl:params}). The particle layer is evident from the local minimum at a distance of one particle diameter from the wall, as shown in the inset where the horizontal axis is scaled with $D_p$. 

The apparent mean particle-to-fluid slip velocity is highest close to the wall and becomes negligible at wall-normal distances $y\gtrsim D_p$, see Fig.~\refsubfig{fig:nmean_out}{b} where we report the wall-normal profiles of the mean particle and fluid velocity for two of the cases considered. Away from the wall, the complex interaction between the turbulent fluid motion and the particles still result in approximately the same average value of streamwise velocity, as if the two phases behave as a continuum. The layer of particles near the wall shows an almost constant slip with respect to the fluid. This large slip indicates that continuum models based on an effective viscosity are bound to fail.

The inset of Fig.~\refsubfig{fig:nmean_out}{b} reports the fluid velocity divided by the particle-to-fluid slip velocity at the wall, $u_{pw}$, versus the wall-normal distance in units of particle radius $R_p$. For the same volume fraction of $20\%$, results from different numerical simulations with different Reynolds numbers and particle sizes collapse for wall-normal distances smaller than a particle radius. 
It appears that, in dense suspensions, a particle-wall layer exists that prevents a direct interaction between the turbulent suspension flow in the core and the solid wall underneath the particle-wall layer. This serves as starting point for the scaling arguments presented hereafter.

%
%%fakesection %SCALING
%
The former considerations motivate a modeling approach based on the separation between the dynamics of the particle-wall layer and of the turbulent flow region. We will denote the latter as the homogeneous suspension region (HSR), meaning a well mixed suspension. Let us therefore define the thickness of the particle-wall layer by the length scale $\delta_{pw}$. The previous discussion showed that $\delta_{pw}$ scales with $D_p$ at fixed volume fraction. In addition, $\delta_{pw}$ should vanish in the single-phase limit, i.e.\  when $\Phi \to 0$. We therefore assume $\delta_{pw}$ (i) to be proportional to the \emph{solidity} of the bulk suspension, measured as the ratio between particle size and mean particle separation distance, and (ii) to scale with the particle size. These hypotheses give the result above $\delta_{pw}=C\,(\Phi/\Phi_{max})^{1/3}\, D_p$, where the constant is set to $C = 1.5$~\footnote{Value obtained by fitting the agreement of the drag to the  simulation results.} for all the cases addressed here and $\Phi_{max} = 0.6$. Note that displacing the origin of the turbulent region has been successfully adopted in turbulent flows over rough walls~\cite{Jackson-JFM-1981}, but was not applied before to the case of turbulent suspensions. \par
\begin{figure}
  \input{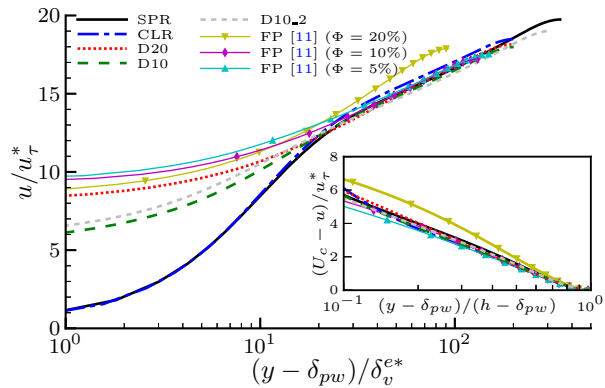}
  \caption{(color online). Profiles of mean streamwise fluid velocity $u/u_\tau^*$ versus the wall-normal coordinate $(y-\delta_{pw})/\delta_v^{e*}$; The inset shows the defect law, $(U_c-u)/u_\tau^*$, versus the distance to the wall in outer units $(y-\delta_{pw})/(h-\delta_{pw})$ (definitions in the text). Maximum statistical error is the same as in Fig.~\ref{fig:umean_fluid_inn}.\label{fig:umean_fluid_inn_correc}}
\end{figure}
\begin{figure}
  \input{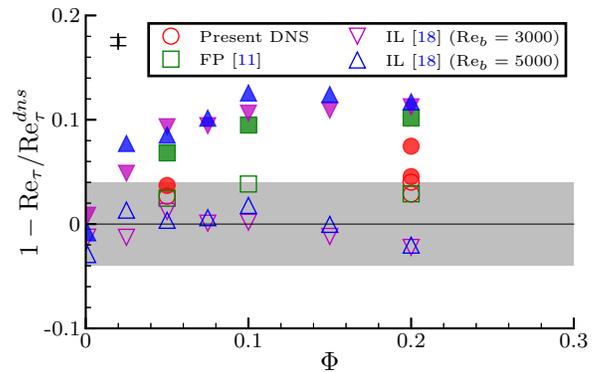}
  \caption{(color online). Relative difference to the theoretical prediction of friction Reynolds number (the shaded area corresponds to a difference of $\pm4\%$). Filled symbols correspond to values that were not for corrected for the presence of the particle-wall layer (i.e.\ $\delta_{pw}=0$). The maximum statistical error in the computation of the overall drag (with $95\%$ confidence interval) from the DNS is below $1\%$. The corresponding (shifted) error bar is also shown on the left hand side of the plot legend.\label{fig:reynolds_compare_err}}
\end{figure}

In the same spirit, we further assume that the total stress $\tau=\rho u_\tau^2(1-y/h)$ acting across the channel is due to two distinct mechanisms. In the HSR, the increment in stress due to the particles is assumed to be well modeled by an effective suspension viscosity; in the particle-wall layer, instead, the stress increases due to the large apparent slip velocity near the wall. This is the main finite-size effect present in the flow.
The stress in the HSR ($y>\delta_{pw}$) corresponds therefore to that of a single-phase turbulent flow of a Newtonian fluid with viscosity $\nu^e$, in a channel with a wall origin at $y=\delta_{pw}$ and half-height $h-\delta_{pw}$. The flow in this region experiences an apparent stress $\rho u_\tau^{*2}\le\rho u_\tau^2$. In the particle-wall layer ($y<\delta_{pw}$) the stress increases linearly when approaching the wall from $\rho u_\tau^{*2}$ to $\rho u _\tau^2 = \rho u_\tau^{*2} + \Delta \tau_{pw}$. Hence, the total stress, linearly varying across the channel~\cite{Picano-et-al-JFM-2015}, is split into two contributions:
\begin{align}
\tau = &\left(\rho u_\tau^{*2}+\Delta \tau_{pw}(1 -   y/\delta_{pw})\right)\mathcal{H}(\delta_{pw}-y) \nonumber \\
     + &\left(\rho u_\tau^{*2}                 (h-y)/(h-\delta_{pw})\right)\mathcal{H}(y-\delta_{pw})\mathrm{;}
\label{eqn:stress_explain}
\end{align}
where $\mathcal{H}$ is the Heaviside step function with the half-maximum convention. Evaluating Eq.~\eqref{eqn:stress_explain} at $y=\delta_{pw}$ yields the friction velocity in this region $u_\tau^* = u_\tau(1-\delta_{pw}/h)^{1/2}$. Given $u_\tau^*$, $\nu^e$ and $\delta_{pw}$ we obtain the following laws for the inner- ($u/u_\tau^*=F[ (y-\delta_{pw}) u_\tau^*/\nu^e]$ ) and outer-scaling ($(U_c-u)/u_\tau^*= G[(y-\delta_{pw})/(h-\delta_{pw})]$) of the mean velocity in the overlap region of the HSR:
%Our results can be summarized as follows. (i) The overall drag of the turbulent suspension is always higher than what predicted by only accounting for the effective suspension viscosity. This increase is attributed to a near-wall layer of particles, denoted as the \emph{particle-wall layer}. (ii) We determine the effective thickness of the particle-wall layer as $\delta_{pw}=C\,(\Phi/\Phi_{max})^{1/3}\, D_p$, with $C = O(1)$ and $\Phi_{max}\approx 0.6$ the maximum packing fraction. (iii) Away from the particle-wall layer the mean flow is well described by a Newtonian fluid with viscosity $\nu^e$, obtained by known empirical laws for suspensions, here $\nu^e/\nu = (1+(5/4)\Phi/(1-\Phi/\Phi_{max}))^2$ \cite{Stickel-and-Powell-ARFM-2005}.  
%The mean velocity and velocity defect in the overlap region become
\begin{align}
  \frac{u}{u_\tau^*} = &\frac{1}{\kappa} \ln\left(\frac{y-\delta_{pw}}{\delta_v^{e*}}\right) + B \mathrm{,} \label{eqn:scaling_um} \\
  \frac{U_c - u}{u_\tau^*} = - &\frac{1}{\kappa} \ln\left(\frac{y-\delta_{pw}}{h-\delta_{pw}}\right) + B_d \mathrm{,} \label{eqn:scaling_ud}
\end{align}
with $u_\tau^* = u_\tau(1-\delta_{pw}/h)^{1/2}$, $\delta_v^{e*} = \nu^e/u_\tau^*$; $\kappa$, $B$ and $B_d$ retain the values of single-phase flow; here $\nu^e/\nu = (1+(5/4)\Phi/(1-\Phi/\Phi_{max}))^2$ \cite{Stickel-and-Powell-ARFM-2005}.
%(iv) The overall drag, i.e.\ the mean wall shear stress $\tau_w=\rho u_\tau^2$, is given by Eqs.~\eqref{eqn:scaling_um} and \eqref{eqn:scaling_ud}, and reads in terms of the friction Reynolds number
% \begin{equation} \mathrm{Re}_\tau =
%   \frac{\mathrm{Re}_b}{2\xi_{pw}^{1/2}}\left(\frac{1}{\kappa}\left[\ln\left(\mathrm{Re}_\tau\chi^e\xi_{pw}^{3/2}\right)-1\right]+B+B_d\right)^{-1}\mathrm{,}
% \label{eqn:scaling_ret}
%\end{equation}
%where $\xi_{pw} = (1-\delta_{pw}/h)$ and $\chi^e=\nu/\nu^e$. Note that Eq.~\eqref{eqn:scaling_ret} reduces to the well-known relation for single-phase flow when $\Phi\to 0$.
%
Fig.~\ref{fig:umean_fluid_inn_correc} reports the mean velocity profiles from the present simulations and the cases from~\cite{Picano-et-al-JFM-2015}. The figure shows a collapse of the profiles in the logarithmic region, except for the case FP \cite{Picano-et-al-JFM-2015} with $\Phi=20\%$ (see Table~\ref{tbl:params}). This is expected from our model, because it is the only case for which the friction Reynolds number based on the scaling parameters of the HSR $\mathrm{Re}_\tau^{hsr} = (h-\delta_{pw})/\delta_v^{e*}<100$. This implies that there is not a sufficient separation of the inner and outer scales for the overlap region to exist \cite{Pope2000}, which is a necessary condition for the logarithmic scaling of the velocity profile. The defect law is shown in outer scaling in the inset of Fig.~\ref{fig:umean_fluid_inn_correc}, where scaling in the logarithmic region can be clearly depicted. Also for this quantity the improvement with respect to the case where the particle-wall layer is not considered ($\delta_{pw}=0$) is significant (not shown).
Finally, the proposed scaling laws are used to derive the following drag law (i.e.\ the mean wall shear stress $\tau_w=\rho u_\tau^2$), expressed in terms of the friction Reynolds number:
 \begin{equation} \mathrm{Re}_\tau =
   \frac{\mathrm{Re}_b}{2\xi_{pw}^{1/2}}\left(\frac{1}{\kappa}\left[\ln\left(\mathrm{Re}_\tau\chi^e\xi_{pw}^{3/2}\right)-1\right]+B+B_d\right)^{-1}\mathrm{,}
 \label{eqn:scaling_ret}
\end{equation}
where $\xi_{pw} = (1-\delta_{pw}/h)$ and $\chi^e=\nu/\nu^e$. Eq.~\eqref{eqn:scaling_ret} is derived in the same way as well-known laws from single-phase flow are derived~\cite{Dean1978}: by integrating the defect law (Eq.~\eqref{eqn:scaling_ud}) over the entire HSR to relate the bulk and centerline velocities, and combining Eqs.~\eqref{eqn:scaling_um} and \eqref{eqn:scaling_ud} to relate the friction and bulk velocities, see supplementary material.  Note that Eq.~\eqref{eqn:scaling_ret} reduces to the well-known relation for single-phase flow when $\Phi\to 0$.
Fig.~\ref{fig:reynolds_compare_err} compares the relative difference between the predicted values of $\mathrm{Re}_\tau$ and the values obtained from the DNS, $\mathrm{Re}_\tau^{dns}$. The filled symbols correspond to predictions where only the effective viscosity is taken into account, i.e.\ $\delta_{pw}=0$, and the open symbols to predictions where both effects are accounted for. The estimates of the drag improve for the three datasets and the difference with the DNS values is less than $4\%$. This supports the necessity of accounting for finite-size effects and further validates the proposed scaling. 
 %
% {\color{red}  Note that the scaling proposed here is based on Prandtl theory valid in the limit of high Reynolds numbers. 
%Empirical relations can be instead used to relate $Re_\tau$ and $Re_b$ at the lower Reynolds numbers accessible to DNS, which give a generally better collapse of the data at high $\Phi$ (lower $Re_\tau$), see supplementary material.}
% Note that the scaling proposed here is based on Prandtl theory valid for sufficiently high Reynolds numbers. Empirical relations can be used instead to relate $\mathrm{Re}_\tau$ and $\mathrm{Re}_b$ at lower Reynolds numbers (easier to reach by DNS), which yield a generally better prediction of the drag at high $\Phi$ and $D_p$ (lower $\mathrm{Re}_\tau^{hsr}$), see supplementary material.
We remark that the implicit formulation of the drag law given by Eq.~\eqref{eqn:scaling_ret} can be replaced by a simple explicit power law of $\mathrm{Re}_\tau$ as a function of $\mathrm{Re}_b$, less sensitive to insufficient inner-to-outer scale separation, which yields similar (and consistently, slightly more accurate at low Reynolds numbers) predictions for the drag, see supplementary material.
  
 %
%Fig.~\refsubfig{fig:reynolds_compare_err}{b} displays the 
The solution of Eq.~\eqref{eqn:scaling_ret}, normalized with the corresponding friction Reynolds number for single-phase flow $\mathrm{Re}_\tau^{sph}={\mathrm{Re}_\tau}|_{\Phi=0}$, can be examined to draw general conclusions on the suspension behavior. 
For constant volume fraction and Reynolds number we conclude that a finite particle size causes a significant increase in drag with respect to the continuum limit due to the formation of a particle-wall layer.
%Assuming constant volume fraction and Reynolds number we confirm that the finite-size particles cause a significant increase of the drag with respect to the continuum limit. This is due to the formation of the particle-wall layer. 
%
As expected, the drag increases monotonically with the particle size (corresponding to an increase of $\delta_{pw}$) and volume fraction (increasing $\delta_{pw}$ and $\nu^e$).
 % Increasing the Reynolds number at fixed particle-size and volume fraction has little effect on the drag 
 %(as expected from the almost-linear dependency $\mathrm{Re}_\tau \sim \mathrm{Re}_b^{0.88}$ \cite{Pope2000}).
 %, except for a combination of very large particles with volume fractions reaching $\Phi=30\%$ and low Reynolds numbers. 
 %This region 

% Clearly, the solution at high values of $\Phi$ and $D_p/h$, and low $Re_b$ is outside the range of validity of our model, as extreme particle confinement and  high effective viscosity invalidate the underlying assumptions of scale separation. 

To conclude, we presented scaling laws for the mean velocity and the velocity defect in turbulent channel flow of neutrally-buoyant finite-size spherical particles, which also enables us to accurately predict the total suspension drag. 
The model quantifies the main finite-size effect present in the flow -- a particle-wall layer which always causes an increase in drag, by separating the dynamics of the 
flow in this layer and the homogeneous suspension region in the core. 
Exploiting conservation of  momentum,
 this effect can be reduced to an
%is accounted for through an 
apparent wall location $y=\delta_{pw}$ above which the flow is reasonably well represented by a Newtonian fluid with an effective suspension viscosity $\nu^e$. 
We validated our predictions for a reasonably wide range of the governing parameters.\par 

%NO The next step is to scrutinize and extend it further against different geometries or boundary conditions, such as pipe flows and boundary layers, both numerically and experimentally.
%%fakesection %ACKNOWLEDGEMENTS
\begin{acknowledgments}
This work was supported by the Portuguese Foundation for Science and Technology under the grant No.\ SFRH/BD/85501/2012, by the European Research Council Grant No.\ ERC-2013-CoG-616186, TRITOS, by the Swedish Research Council (VR) and by COST Action MP1305: Flowing Matter. We acknowledge computer time provided by SNIC (Swedish National Infrastructure for Computing) and PRACE project 2014112543 for awarding us access to resource CURIE based in France at Genci/CEA. Iman Lashgari is thanked for kindly providing the data from \cite{Lashgari-et-al-PRL-2014} in digital form.
\end{acknowledgments}
%%fakesection %BIBLIOGRAPHY
\bibliography{bibfile}

\clearpage
\renewcommand\thefigure{S\arabic{figure}}
\setcounter{figure}{0}
\renewcommand{\theequation}{S\arabic{equation}}
\setcounter{equation}{0}

\section{Derivation of Drag Law}

%As done for single-phase turbulent channel flow, 
We derive a relation for the frictional drag, expressed in terms of a friction Reynolds number, $\mathrm{Re}_\tau$, from the scaling considerations presented in the Letter. The nomenclature is consistent with the one used in the Letter. In the single-phase case, one can  relate the bulk velocity to the friction velocity from the logarithmic scaling laws for mean velocity and velocity defect. A detailed derivation, together with the inherent assumptions can be found e.g.\ in~[S.\ B.\ Pope \textit{Turbulent flows}. Cambridge university press, 2000].\par
We aim at relating $\mathrm{Re}_\tau$ to the parameters governing the flow in the overlap region: $\mathrm{Re}_b$, $\Phi$ and $D_p/h$. 
As for
% Here we use assumptions analogous to those that are used in 
single-phase turbulent channel flow we assume for the homogeneous suspension region that the bulk velocity is well approximated by integrating the velocity defect over the height of the homogeneous suspension region (HSR). This approximation is valid as long as (i) the Reynolds number is sufficiently high that the inner layer of the HSR does not contribute significantly to the bulk velocity and (ii) the virtual wall origin $\delta_{pw}$ is sufficiently small that the flow inside the particle-wall layer contributes little to the bulk velocity. Thus, 
\begin{align}
U_b &\approx \frac{1}{h-\delta_{pw}}\left(\int_{\delta_{pw}}^{h} u \, \mathrm{d}y \right).
\end{align}
\par
The bulk velocity is then estimated by integrating the defect law from $\delta_{pw}$ to $h$ (consistency requires that the constant $B_d$, typically small, is set to $0$),
\begin{align}
U_b \approx \left(U_c - \frac{u_\tau^*}{\kappa}\right).
\label{eqn:defect_integral}
\end{align}

Next, the two expressions for the log law, in inner and outer variables respectively, are combined to relate the mean centerline velocity $U_c$ to the apparent wall friction velocity $u_\tau^*$, yielding:
\begin{align}
\frac{U_c}{u_\tau^*} = \frac{1}{\kappa}\ln\left(\frac{h-\delta_{pw}}{\delta_v^{e*}}\right) + B + B_d
\label{eqn:combine_defect_mean}
\end{align}
Combining Eqs.~\eqref{eqn:defect_integral}~and~\eqref{eqn:combine_defect_mean} we obtain the following expression for $U_b/u_\tau^*$:
\begin{align}
\frac{U_b}{u_\tau^*} = \frac{1}{\kappa}\left[\ln\left(\frac{h-\delta_{pw}}{\delta_v^{e*}}\right)-1\right]+B+B_d
\label{eqn:volume_flux_homogeneous}
\end{align}

Substituting  $u_\tau^* = u_\tau(1-\delta_{pw}/h)^{1/2}$, and $\delta_v^{e*} = \nu_e/u_\tau^*$ in Eq.~\eqref{eqn:volume_flux_homogeneous} we get
\begin{widetext}
\begin{align}
\frac{U_b}{u_\tau} = \left(\frac{1}{\kappa}\left[\ln\left(\mathrm{Re}_\tau\frac{\nu}{\nu^e}\left(1-\frac{\delta_{pw}}{h}\right)^{3/2}\right)-1\right]+B+B_d\right)\left(1-\frac{\delta_{pw}}{h}\right)^{1/2}
\label{eqn:volume_flux_homogeneous2}
\end{align}
\end{widetext}
After re-arranging, we finally obtain
\begin{align} 
  \mathrm{Re}_\tau = \frac{\mathrm{Re}_b}{2\xi_{pw}^{1/2}}\left(\frac{1}{\kappa}\left[\ln\left(\mathrm{Re}_\tau\chi^e\xi_{pw}^{3/2}\right)-1\right]+B+B_d\right)^{-1} \mathrm{,}
  \label{eqn:drag_final}
\end{align}
where $\xi_{pw} = (1-\delta_{pw}/h)$ and $\chi^e=\nu/\nu^e$. Eq.~\eqref{eqn:drag_final} can be solved numerically by substituting $\delta_{pw} = C(\Phi/\Phi_{max})^{1/3}D_p$ and $\nu_e=(1+(5/4)\Phi/(1-\Phi/\Phi_{max}))^2\nu$. The constant $C=O(1)$ was set to $1.5$ for all the cases presented in this study, and $\Phi_{max}$ to $0.6$.
\section{An alternative correlation for the overall drag}

\begin{figure}
  \centering
  \includegraphics{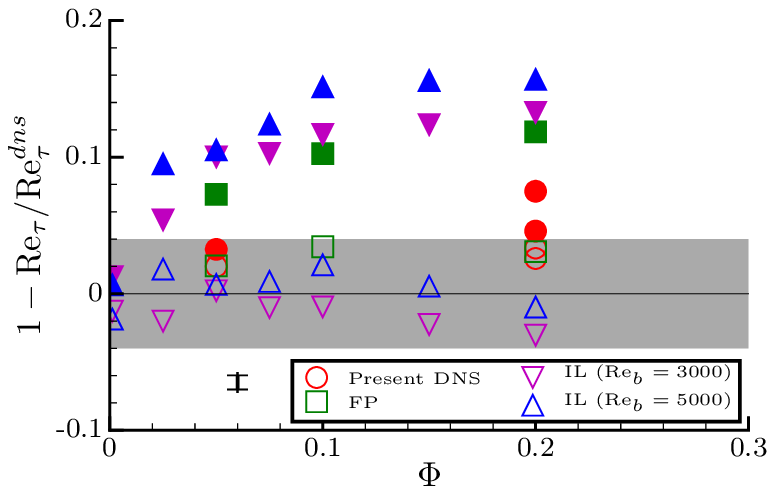}
  \caption{Relative difference to the theoretical prediction of friction Reynolds number (the shaded area corresponds to a difference of $\pm4\%$) when an empirical correlation is used: $\mathrm{Re}_\tau = 0.09(\mathrm{Re}_b\chi^e\xi_{pw})^{0.88}/(\xi^{3/2}\chi^e)$. Filled symbols correspond to values that were not for corrected for the presence of the particle-wall layer (i.e.\ $\delta_{pw}=0$).The maximum statistical error in the computation of the overall drag (with $95\%$ confidence interval) from the DNS is below $1\%$. The corresponding (shifted) error bar is also shown on the left hand side of the plot legend.} 
\label{fig:reynolds_compare_err_empirical}
\end{figure}

Fig.~\ref{fig:reynolds_compare_err_empirical} displays the same quantity as Fig.~4 of the manuscript: the relative difference between predicted values of $\mathrm{Re}_\tau$ and the values obtained from the DNS, $\mathrm{Re}_\tau^{dns}$. 
The difference now is that the estimate is based on an empirical correlation valid for single-phase flow ($\mathrm{Re}_\tau^{sph} \approx 0.09\mathrm{Re}_b^{0.88}$~[S.\ B.\ Pope \textit{Turbulent flows}. Cambridge university press, 2000]), which is extended to the case of a turbulent suspension. 
For the homogeneous suspension region (see the modeling considerations in the Letter) we obtain:
\begin{equation}
  \mathrm{Re}_\tau^{hsr} = \frac{u_\tau^*(h-\delta_{pw})}{\nu^e} \approx 0.09\left(\frac{U_b(h-\delta_{pw})}{\nu^e}\right)^{0.88}\mathrm{,}
\end{equation}
and from this we derive the following explicit, power-law expression for $\mathrm{Re}_\tau$:
\begin{equation}
  \mathrm{Re}_\tau = \frac{0.09\left(\mathrm{Re}_b\chi^e\xi_{pw}\right)^{0.88}}{\xi_{pw}^{3/2}\chi^e}\mathrm{;}
\label{eqn:retau_empirical}
\end{equation}
where $\xi_{pw} = (1-\delta_{pw}/h)$ and $\chi^e=\nu/\nu^e$. 
Fig.~\ref{fig:reynolds_compare_err_empirical} shows that the empirical correlation given by Eq.~\eqref{eqn:retau_empirical} yields similar predictions for the drag as Eq.~(3) in the Letter. In general, the predictions from the empirical correlation are slightly more accurate (i.e., the error is smaller), in particular for the data at the lowest values of $\mathrm{Re}^{hsr}$ (see upward- and downward-pointing triangles).
%The difference between the predicted values and the DNS is lower
%for the data at the lowest values of $\mathrm{Re}^{hsr}$ (uppward- and downward-pointing triangles). 
%The reason is that this empirical relation is based on DNS data at lower Reynolds, and do not assume complete scale separation as Prandtl theory used in Eq.~3 of the manuscript. Expression~\eqref{eqn:retau_empirical} is therefore a valid alternative to Eq.~3 in the manuscript, especially at moderate Reynolds numbers.

It is interesting to note that the explicit nature of Eq.~\eqref{eqn:retau_empirical} enables us to estimate the relative importance of the finite-size effect ($\xi_{pw}$) and effective suspension viscosity ($\chi^e$) at the given flow rate (quantified by $\mathrm{Re}_b$):
\begin{equation}
  \mathrm{Re}_\tau \propto \frac{\mathrm{Re}_b^{0.88}}{\xi_{pw}^{0.62}\chi^{e\,0.12}}\mathrm{.}
\label{eqn:retau_empirical2}
\end{equation}
The large exponent of $\xi_{pw}$, $0.62$, confirms that the finite-size effect plays an important role. %To illustrate this, we compute from Eq.~\eqref{eqn:retau_empirical2} that for case D10 the finite-size effect alone contributes to $1.8\%$ to the increase in the (single-phase flow) drag and the effective viscosity effect alone to $7.9\%$ increase in the drag. For case D20, for which the effective viscosity is the same as D10, the sole finite-size effect yields an increase of $3.8\%$ in the drag, which is more than twice as large as the finite-size contribution in case D10 and equal to nearly half of the effective-viscosity contribution.
%Since $(\nu_e-\nu)/\nu \sim \Phi + O(\Phi^2)$ increases faster with $\Phi$ than $\delta_{pw}/h \sim \Phi^{1/3}D_p/h$, it follows from Eq.~\eqref{eqn:retau_empirical2} that the relative importance of the finite-size effect becomes more pronounced when lowering $\Phi$ and increasing $D_p/h$.
At fixed particle size, however, the effective viscosity still plays a major role in these dense flows, as $1/\chi^e \sim 1 + \Phi + O(\Phi^2)$ increases faster with $\Phi$ than $1/\xi_{pw}\sim 1+\Phi^{1/3}+O(\Phi^{2/3})$.
\end{document}